\documentclass[sigplan,screen,authorversion]{acmart}
\pdfoutput=1

\usepackage[utf8]{inputenc}
\usepackage[T1]{fontenc}

\usepackage{microtype}
\usepackage{listings}
\title{CHERI Performance Enhancement for a Bytecode Interpreter}
\author{Duncan Lowther}
\affiliation{%
    \institution{University of Glasgow}
    \city{Glasgow}
    \country{United Kingdom}
}
\email{duncan.lowther@glasgow.ac.uk}
\orcid{0009-0004-9310-8092}

\author{Dejice Jacob}
\affiliation{%
  \institution{University of Glasgow}
  \city{Glasgow}
  \country{United Kingdom}
}
\email{dejice.jacob@glasgow.ac.uk}
\orcid{0000-0002-4137-0353}

\author{Jeremy Singer}
\affiliation{%
    \institution{University of Glasgow}
    \city{Glasgow}
    \country{United Kingdom}
}
\email{jeremy.singer@glasgow.ac.uk}
\orcid{0000-0001-9462-6802}

\setcopyright{acmlicensed}
\acmPrice{15.00}
\acmDOI{10.1145/3623507.3623552}
\acmYear{2023}
\copyrightyear{2023}
\acmSubmissionID{splashws23vmilmain-p65-p}
\acmISBN{979-8-4007-0401-7/23/10}
\acmConference[VMIL '23]{Proceedings of the 15th ACM SIGPLAN International Workshop on Virtual Machines and Intermediate Languages}{October 23, 2023}{Cascais, Portugal}
\acmBooktitle{Proceedings of the 15th ACM SIGPLAN International Workshop on Virtual Machines and Intermediate Languages (VMIL '23), October 23, 2023, Cascais, Portugal}
\received{2023-07-23}
\received[accepted]{2023-08-28}

\begin{document}

\begin{abstract}

During our port of the MicroPython bytecode interpreter to the CHERI-based Arm Morello platform, we encountered a number of serious performance degradations. This paper explores several of these performance issues in detail, in each case
we characterize the cause of the problem, the fix, and the corresponding interpreter performance improvement over a set of standard Python benchmarks.

While we recognize that Morello is a prototypical physical instantiation of the CHERI concept, we show that it is possible to eliminate certain kinds of software-induced runtime overhead that occur due to the larger size of CHERI capabilities (128 bits) relative to native pointers (generally 64 bits). 
In our case, we reduce a geometric mean benchmark slowdown from 5x (before optimization) to 1.7x (after optimization) relative to AArch64, non-capability, execution. The worst-case slowdowns are greatly improved, from 100x (before optimization) to 2x (after optimization).

The key insight is that implicit pointer size presuppositions pervade systems code; whereas previous CHERI porting projects highlighted compile-time and execution-time errors exposed by pointer size assumptions, we instead focus on the performance implications of such assumptions.

\end{abstract}



\maketitle

\section{Introduction}
\label{sec:intro}

The CHERI concept of microarchitectural capabilities involves
processor support for fat pointers and hardware checks on memory accesses.
This radical, perhaps invasive, approach to improve dynamic memory safety
is massively challenging for programming language runtimes.
In previous work \cite{singer23towards,lowther2023mplr}, we described our strategy for
adapting the MicroPython runtime to run on a CHERI platform.
In this earlier work, we tackled two key families of bugs:
\begin{enumerate}
  \item CHERI-induced compiler errors, caused by pointer size assumptions in the code base, or legacy C style pointer abuse; and,
  \item CHERI-induced runtime errors, due to capability violations in
    memory accesses---either out-of-bounds accesses or invalid capabilities.
\end{enumerate}

Fixing these classes of bugs enabled us to get MicroPython up-and-running on
CHERI; however the initial performance is poor. There appear to be significant runtime overheads and inefficiencies associated with CHERI.

In this work, we look at a further crucial stage in porting
existing virtual machines
to CHERI: i.e.\ \textit{performance optimization}.
Suppose we can get something running on a CHERI platform, how do we
ensure it runs at an acceptable speed?

This paper starts with a working, but highly inefficient, port of MicroPython
to CHERI; this version has a geometric mean slowdown of 5x (relative to equivalent non-CHERI execution) for a set of standard Python benchmarks.  We follow a performance debugging strategy to identify and eliminate
runtime overheads, noting these are mainly associated with dynamic memory management. After three rounds of software performance optimization, our CHERI MicroPython has a geometric mean slowdown of 1.7x (relative to non-CHERI). We expect there are further performance gains to be made, but this work clearly demonstrates that careful performance profiling and debugging is essential if CHERI is
to be adopted by the language runtime community.

\section{Background}
\label{sec:bg}


\subsection{CHERI and Morello}
A Capability Hardware Enhanced RISC Instructions (CHERI) system
\cite{woodruff2014cheri,watson2020capability}
is a collection of instruction set
extensions and processor logic for modern micro-architectures, providing
direct support for embedding metadata into `fat' pointers to enable runtime
checks on the use of these values (which are known as capabilities).
Key properties enforced by CHERI include the following:
\begin{enumerate}
\item Capabilities cannot be forged and have to be derived from an existing capability enforced by a validity tag (so, no casting from \texttt{int} to \texttt{void *}).
\item Capability spatial bounds are tightly enforced, and bounds cannot be `grown', only monotonically reduced.
\item Capabilities have permissions, similar to page-table permissions (\texttt{r/w/x}) but they enable fine-grained control -- effectively capabilities provide per-pointer permissions.
\end{enumerate}

The CHERI concept has been instantiated by Arm in the Morello prototype architecture \cite{morello2021}. Morello is a quad-core 64-bit Arm processor (ISA v8.2-A) based on the commercially available Neoverse N1 system. While we recognize there may be minor performance anomalies and inconsistencies in this prototype implementation \cite{watson23arm}, it is a fully-functional platform and can provide useful performance forecasts for CHERI adopters.
Whereas in conventional AArch64 processors, a pointer is simply a 64-bit machine word, on Morello an architectural capability is a 128-bit value that includes an address, bounds, and associated metadata \cite{woodruff2019cheri}. Further, the validity `tag' bit is stored out-of-band and cannot be manipulated directly by user code.

\subsection{CHERI Porting}
A great deal of open-source software has been ported to CHERI and Morello \cite{watson15cheri,davis2019cheriabi}. In particular, the FreeBSD OS has a port named CheriBSD. Many user-space applications have been adapted for CHERI, often with minimal source code changes. A port of the KDE desktop framework reportedly incurred only 0.026\% lines of altered code \cite{watson2021assessing}.

However, systems level code is more likely to feature pointer-intensive operations and unusual interactions with memory. These are the areas where adaptation for CHERI is more complicated. One study of CHERI memory allocators \cite{bramley2023picking} reveals that, for some real-world C library \texttt{malloc} implementations, up to 10\% of the code base requires modification. Further, performance of memory allocators on Morello is inconsistent and the performance profiling tools are not sufficiently mature to diagnose the root causes of problems.

There are a few VM ports to CHERI at least partially underway. Of these, the most complete appears to be a JavaScriptCore port \cite{gutstein2022memory}. However so far, no meaningful performance results are publicly available for this or any other VM on Morello.

\subsection{MicroPython}
MicroPython \cite{micropython} is a small-scale Python interpreter, largely written in C, explicitly targeting microcontroller scale devices.
It is a straightforward bytecode interpreter with a fixed size heap, implementing a non-moving mark/sweep garbage collector.
MicroPython features a set of libraries, some of which are specific for the embedded systems domain, others are general purpose.
The interpreter can be compiled and executed as a standalone, user-space process VM on a POSIX host environment. We have adapted MicroPython for CHERI, based on this process VM model, running on CheriBSD.
A work-in-progress report describes our initial work \cite{singer23towards}
and a follow-on paper describes the complete port \cite{lowther2023mplr}.
Since then, we have been looking at performance optimization, particularly focusing on high-overhead memory management aspects of the VM.

MicroPython comes with a set of benchmarks, some of which are intended for performance measurements. 
These are based in part on the Programming Language Shootout benchmarks \cite{marr16cross},
which are familiar to VM developers across multiple languages. 
In our performance optimization work, as in other assessments of CHERI performance, we compare AArch64 code running on Morello (known as \texttt{hybrid mode}) with capability-enhanced code running on the same platform (known as \texttt{purecap mode}). We measure and report the performance of the purecap code for each benchmark, relative to the equivalent execution of hybrid code. In our case, these are identical bytecode benchmarks running on distinct MicroPython interpreter instances (an AArch64 executable versus a CHERI executable).

Subsequent sections in this paper explore the software performance optimizations we applied for MicroPython on the Morello platform.

\section{Heap Block Size}
\label{sec:block}

\begin{figure*}[ht]
  \begin{center}
    \includegraphics[width=\textwidth]{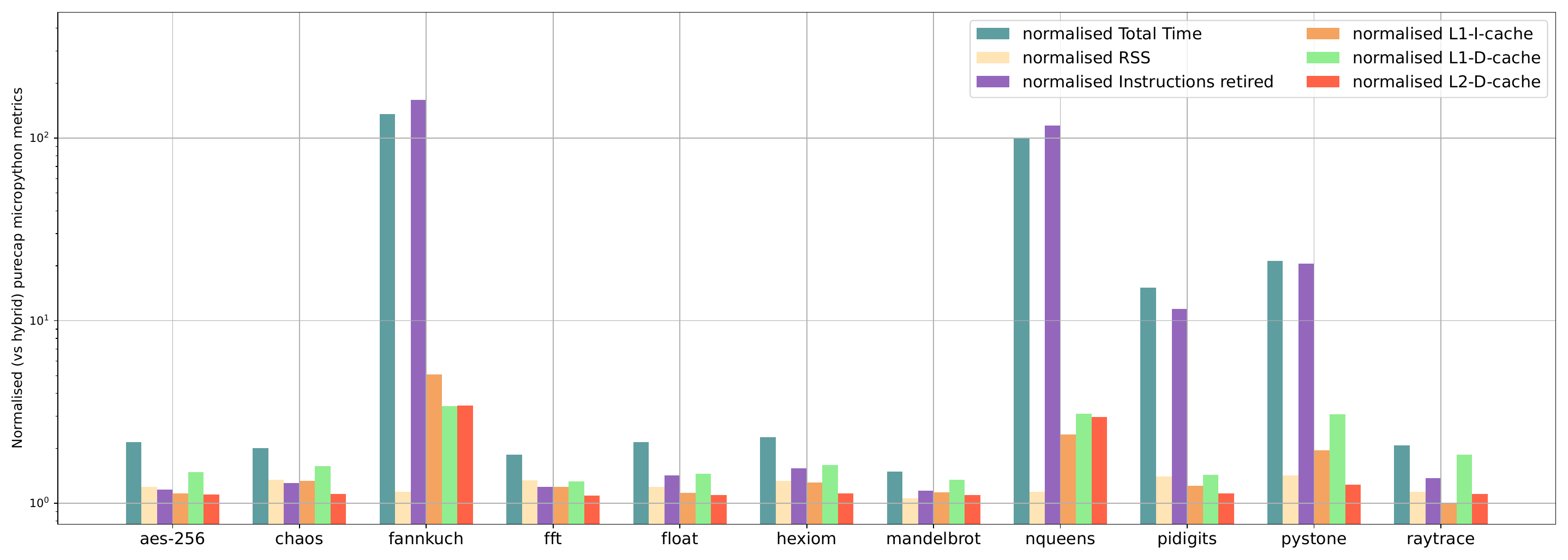}
	  \Description{A logarithmic bar chart showing purecap MicroPython performance metrics, normalised against hybrid performance, for eleven benchmarks. Normalised total time and instructions retired exceed 100x for benchmarks fannkuch and nqueens and are between 10x and 30x for benchmarks pidigits and pystone. All metrics on other benchmarks, vary between $1.0$ and $2.5$.}
  \end{center}
  \caption{\label{fig:initial}
  Performance of Python benchmarks running on the purecap interpreter, normalised
  to the hybrid interpreter performance. For example, the wall-clock
  execution time \emph{total-time} of
  chaos on purecap is 2.0x greater than on hybrid. 
  To understand why purecap is slower than we expected, we recorded
  several performance metrics: \emph{RSS} is
  maximum memory utilisation; \emph{INST\_RETIRED} the number of instructions
  retired while executing the benchmark; and \emph{\{L1-I, L1-D, L2-D\}\_CACHE}
  the L1 instruction, L1 data, and L2 data, cache misses respectively.
  The normalised \textit{y}--axis is on a logarithmic scale.
  }
\end{figure*}

\begin{figure*}[ht]
  \begin{center}
    \includegraphics[width=\textwidth]{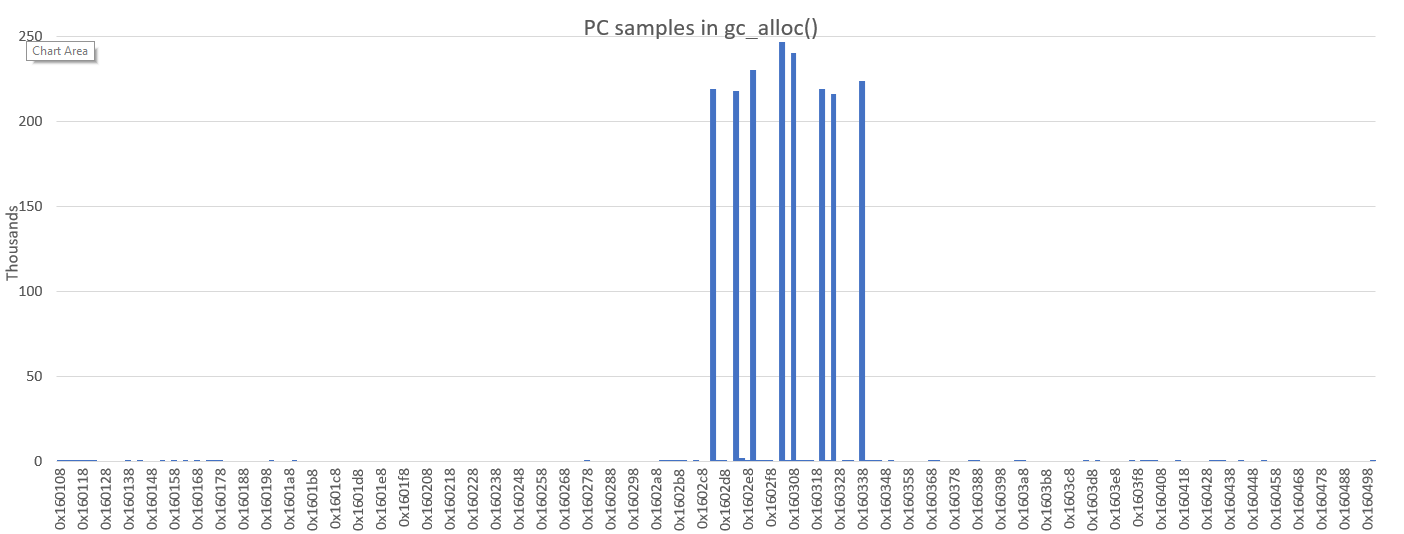}
	  \Description{A histogram showing the distribution of sampled PC values. Between 0x1602d0 and 0x160338 there are 8 addresses which exceed 200,000 samples each; the bars for other addresses are barely visible (<5000). }
  \end{center}
	\caption{\label{fig:pchist} Histogram of sampled PC values from \texttt{gc\_alloc()}}
\end{figure*}

When we first evaluated the performance of our purecap build of the MicroPython interpreter, we saw highly concerning runtime overheads on four of the benchmarks Figure~\ref{fig:initial}. On two of the benchmarks, the purecap execution time exceeded 100x that of the hybrid version (\verb`fannkuch` at 135.7 and \verb`nqueens` at 100.0); two more had overheads above 10x (\verb`pystone` at 21.3 and \verb`pydigits` at 15.2).

We began by tackling the 100x overheads. One thing that \verb`fannkuch` and \verb`nqueens` had in common was repetitive list-slicing operations, and a further list-slicing microbenchmark confirmed that the overhead was caused by those operations. It is also useful to note that the normalised instruction-retired counts were elevated at similar levels to the normalised execution time.

To build a more detailed execution profile, the \verb`pmcstat` utility (a FreeBSD profiling tool) was used
to sample the callchain of the \verb`fannkuch` benchmark at every 65536 instructions retired. The current version of \verb`pmcstat` on CheriBSD had issues resolving the symbols of the purecap binary
By creating a simple program to read the \verb`pmcstat` dump, print out raw memory addresses and performing a manual lookup,
we concluded that 99.3\% (1820989/1833685) of the samples occured in the \verb`gc_alloc()` function in the top stack frame.
The \verb`gc_alloc()` function is an API-level entry point of the heap allocator in MicroPython.

Figure~\ref{fig:pchist} is a histogram of the sampled \texttt{PC} values within the \texttt{gc\_alloc()} function. 
The samples are very clearly concentrated in the interval between \texttt{0x1602d0} and \texttt{0x160338}.
By debugging (using \texttt{gdb}) and mapping these addresses to symbols in the MicroPython binary
pointed to a \texttt{for}-loop in the allocator, used to search for a contiguous run of free blocks of the appropriate length. 
From this, we conjectured that the issue was due to increased memory fragmentation, likely caused by the 
width of a CHERI capability being double the width of a AArch64 pointer

\begin{figure*}[ht]
  \begin{center}
    \includegraphics[width=\textwidth]{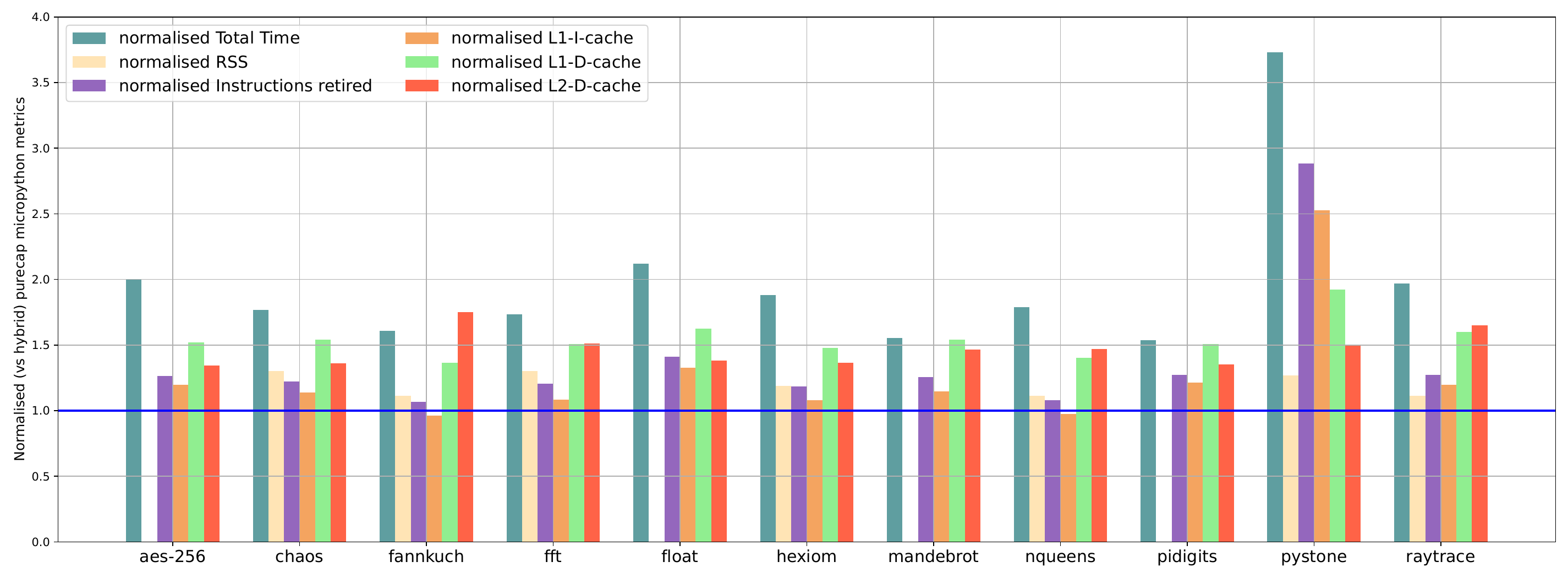}
	  \Description{bar cha showing purecap MicroPython performance metrics, normalised against hybrid performance, for eleven benchmarks. Normalised total time varies between $1.5$ and $2.1$ for most benchmarks, with pystone an outlier at $3.7$. Instructions retired and instruction cache misses vary between $1.0$ and $1.5$ for most benchmarks, with pystone an outlier at $2.9$ instructions retired and $2.5$ cache misses. Data cache misses vary between $1.3$ and $1.9$. }
  \end{center}
  \caption{\label{fig:perf_eval_fixed} Performance of Python benchmarks running on the purecap interpreter with increased block size, normalised to the hybrid interpreter performance.}
\end{figure*}

MicroPython's in-built GC statistics functionality shows a large number of 1-block allocations when 
executing the interpreter compiled in hybrid mode. There were no statistically significant
1-block allocations during execution of the purecap interpreter. 
MicroPython's default blocksize is \texttt{4*sizeof(mp\_uint\_t)},
i.e.\ four times the platform word size.
As both hybrid and purecap use 64-bit integers, they used the same block size, despite purecap objects
being larger due to the 128-bit size of capabilities.
Changing the block size to \texttt{4*sizeof(mp\_obj\_t)} 
(i.e., scaling based on pointer size rather than integer size) restored the
expected frequency of 1-block allocations and fixed the overhead issue, as shown in Figure~\ref{fig:perf_eval_fixed}.

The frequency of 1-block allocations is important, because MicroPython's allocator only advances its `last-free index' (the point at which a new call to \texttt{gc\_alloc()} begins searching for free blocks) when a 1-block allocation is made. This guarantees that there are no free blocks in the `skipped' region, but also means that the index lags significantly behind the actual first free block when most allocations are two or more blocks long. A more sophisticated allocator could update the index more often, but this paper focuses on the removal of purecap-introduced overhead and such changes, being shared with the reference implementation, would be out of scope.

\section{Stack Frame Size}
\label{sec:frame}

\begin{figure*}[ht]
  \begin{center}
    \includegraphics[width=\textwidth]{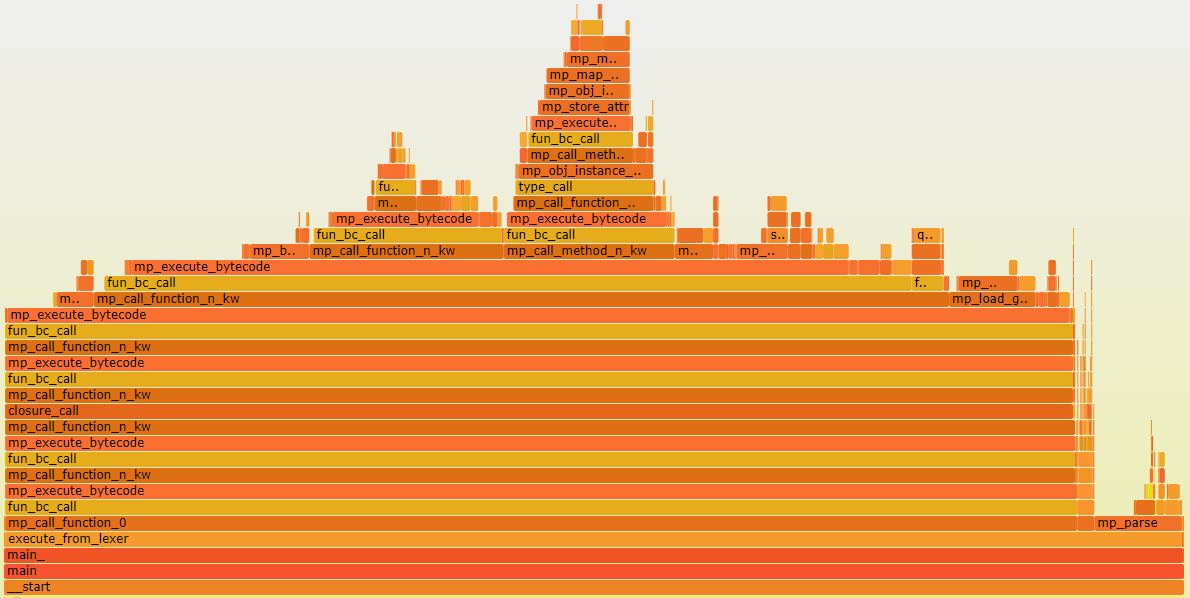}
  \end{center}
  \caption{\label{fig:flame-ainit} 
	Flame graph (in terms of instructions retired) for hybrid version of MicroPython interpreter running \texttt{pystone} benchmark.}
\end{figure*}

\begin{figure*}[ht]
  \begin{center}
    \includegraphics[width=\textwidth]{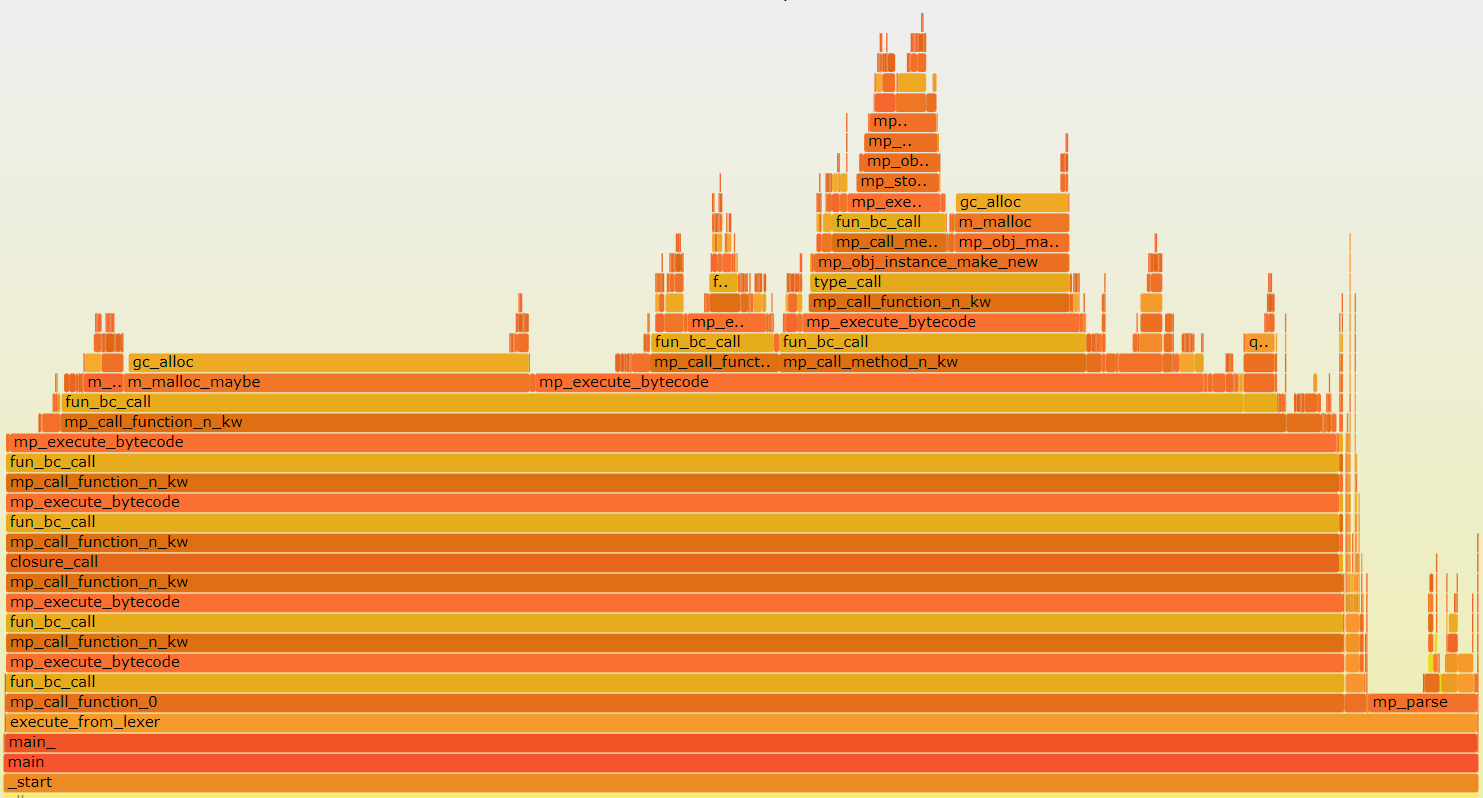}
  \end{center}
  \caption{\label{fig:flame-cinit} 
	Flame graph (in terms of instructions retired) for purecap version of MicroPython interpreter running \texttt{pystone} benchmark---notice the large \texttt{gc\_{}alloc} horizontal bar.}
\end{figure*}

After changing the block-size to a multiple of capability-width, 
the purecap version displayed a execution time, relative to the hybrid version,
of 1.5 to 2.1 times on all benchmarks except \verb`pystone`. \verb`Pystone` executed by
the purecap interpreter showed a normalised slowdown of around 3.7x. 
We thus turned our attention to diagnosing the performance issues in this particular benchmark. 
The custom program that was used to read \verb`pmcstat` dumps to look up addresses in this table was augmented
with the symbol-table extracted from the binary. This is done using the \verb`llvm-objdump-morello` tool. 
We also modified it to output the samples in a format readable by Brendan Gregg's\cite{flamegraphs,flamegraphs2}
Flame Graph scripts.
A \emph{flame graph} shows aggregated stack traces over the full execution of a program,
so the width of each function bar on the graph indicates the amount of time
spent executing that function.
Figures~\ref{fig:flame-ainit} and \ref{fig:flame-cinit} show the flame graphs
of the execution of the \verb`pystone` benchmark, sampled every 65536 instructions. 

From these graphs, it was clear that \verb`gc_alloc()` was again a significant source of overhead.
The \verb`gc_alloc()` allocator function accounted 
for approximately a quarter of the overall instruction count
on the purecap version while being statistically insignificant for the hybrid execution run.
Looking at the program counter samples within \verb`gc_alloc()` again showed a concentration in the search-for-free-blocks loop. Instrumenting the allocator showed that this loop ran over 200 times as many iterations on the purecap version as the hybrid version. Further, there were thousands of 3- and 4- block allocation calls on the purecap version that did not occur on the hybrid version.
Unlike in Section \ref{sec:block}, where a similar symptom was due to the purecap allocations taking up more blocks than their hybrid equivalents, in this case, these were in addition to a comparable number of 1- and 2-block allocations.

Referring back to Figure~\ref{fig:flame-cinit} we see that the most expensive \verb`gc_alloc()` calls are coming from \verb`fun_bc_call()` (through the wrapper \verb`m_malloc_maybe()`). Tracing these calls, we found that they occured when allocating space for the call frame of a Python function. 
The interpreter places smaller stack frames on the C stack and larger stack frames in the heap and is controlled by the configurable compile-time constant  value \verb`VM_MAX_STATE_ON_STACK`.
This constant, like the block size, is calculated based on the integer word size -- a frame of up to 16 words will be allocated on the stack.
On purecap, frames are generally twice as large as hybrid, and so were being 
heap-allocated 
more frequently. 16 words occupy 128 bytes on a 64-bit system, or 2 `new' heap blocks on purecap. Frames that would, on hybrid, fit within this limit may be up to 256 bytes on purecap, or 4 `new' heap blocks. Thus, the frames spilled because of the change to purecap were all 3 or 4 blocks long, resulting in the anomalous increase in allocation calls noted earlier.

\begin{figure*}[ht]
  \begin{center}
    \includegraphics[width=\textwidth]{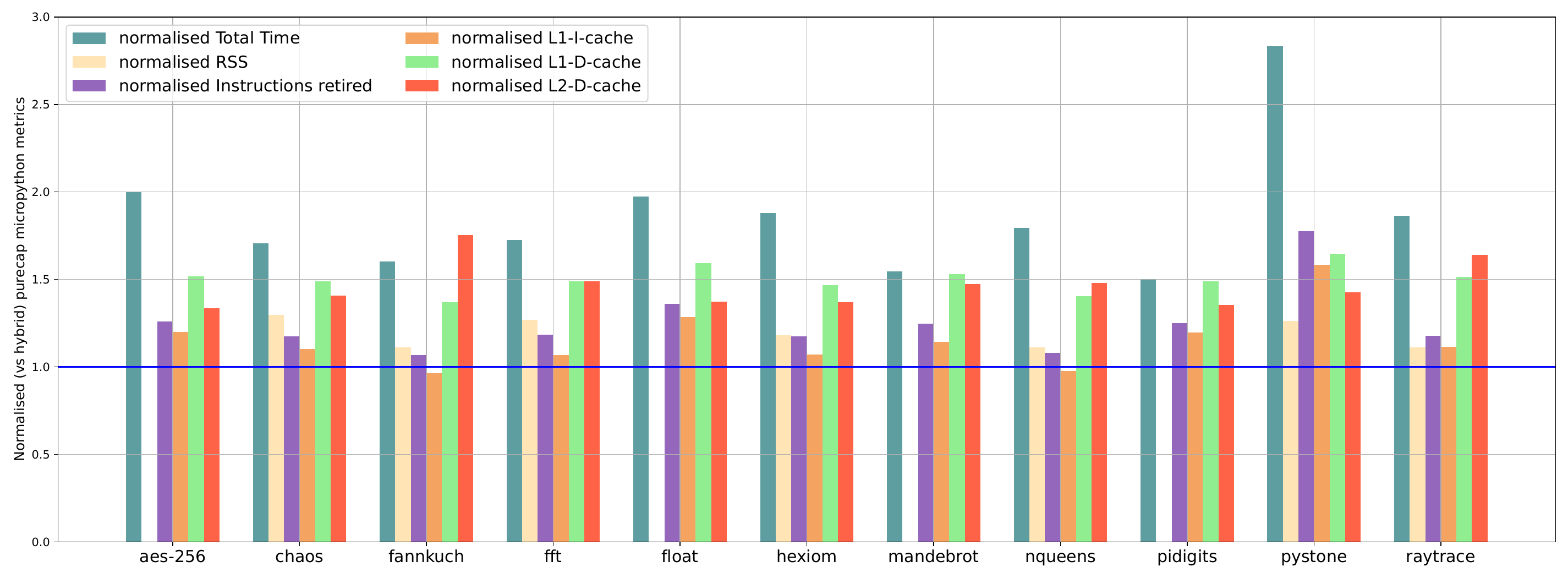}
	  \Description{A bar chart showing purecap MicroPython performance metrics, normalised against hybrid performance, for eleven benchmarks. Normalised total time varies between $1.5$ and $2.0$ and other metrics between $1.0$ and $1.5$ for most benchmarks, with pystone an outlier at $2.8$ normalised total time and >1.5 instructions retired and L1 instruction and L1 data cache misses.}
  \end{center}
  \caption{\label{fig:perf_eval_fixed2} Performance of Python benchmarks running on the purecap interpreter with increased block size and maximum VM state size, normalised to the hybrid interpreter performance.}
\end{figure*}

\begin{figure*}[ht]
  \begin{center}
    \includegraphics[width=\textwidth]{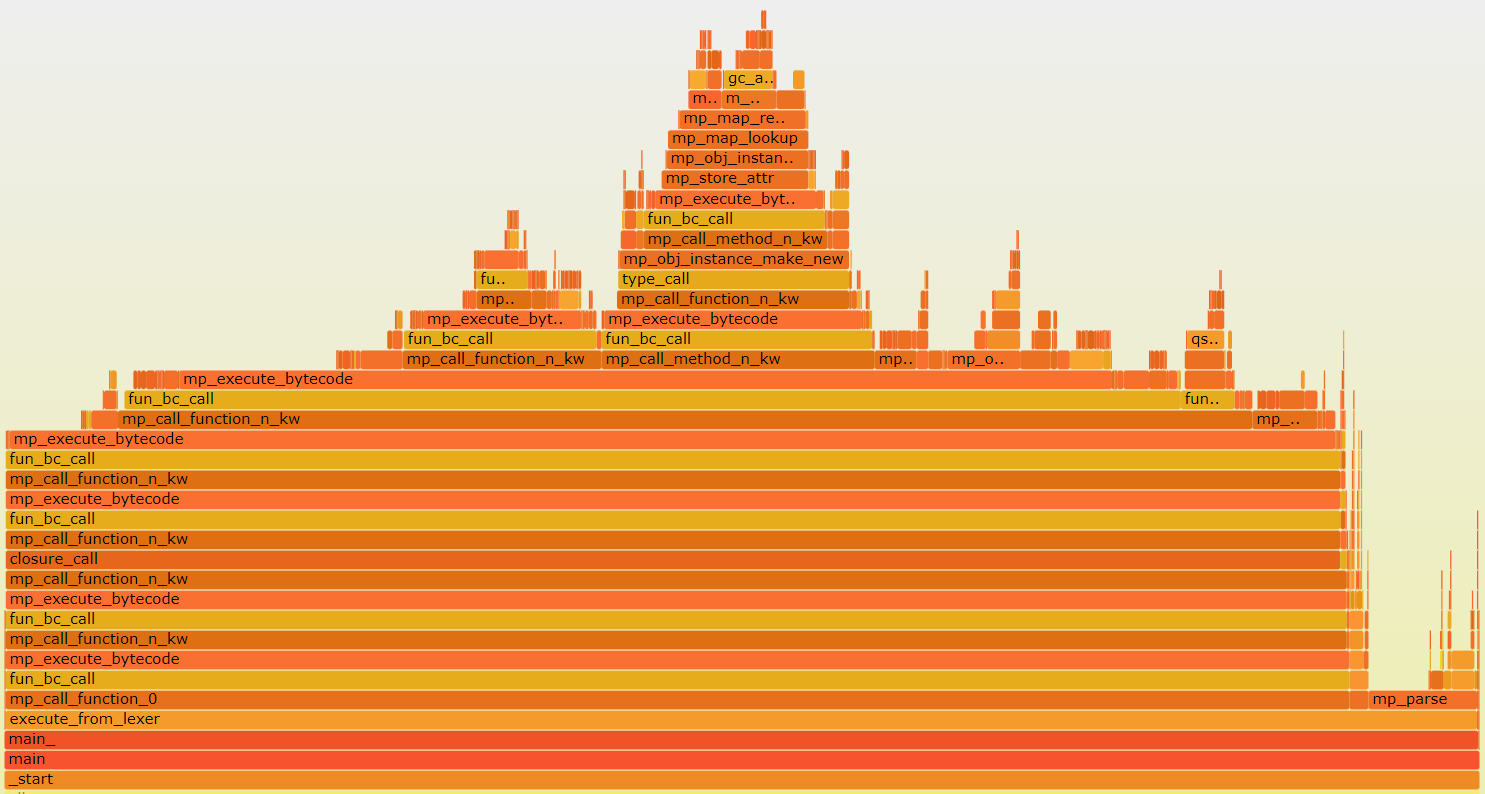}
  \end{center}
  \caption{\label{fig:flame-cstackfix} 
	Flame Graph (instructions retired) of purecap interpreter (after \texttt{VM\_MAX\_STATE\_ON\_STACK} adjustment) running \texttt{pystone} benchmark.}
\end{figure*}

We introduced a change, to redefine the compile-time constant \verb`VM_MAX_STATE_ON_STACK` based on the size of a pointer rather than the size of an integer.
This ensures that call stacks are not spilled where they would be stack-allocated in the reference implementation. Following this adjustment, the purecap interpreter displayed a significant performance improvement, as shown in Figure~\ref{fig:perf_eval_fixed2}. The normalised execution time on the \verb`pystone` benchmark dropped from 3.7 to 2.8, with all other benchmarks under 2.0. The new flame graph (Figure~\ref{fig:flame-cstackfix}) shows that \verb`gc_alloc()` is no longer a significant overhead.

\section{Compilation Inefficiency}
\label{sec:asm}

Following the fixes in the previous two sections, the performance on the \verb`pystone` benchmark, while much improved, was still noticably worse than the other benchmarks. As can be seen in Figure~\ref{fig:perf_eval_fixed2}, while normalised instructions-retired no longer tightly tracks normalised execution time, it is still (at 1.8) a significant contributor to the overhead. We thus once more examine why so many more instructions are being executed by the purecap version.

\begin{figure*}[ht]
  \begin{center}
    \includegraphics[width=\textwidth]{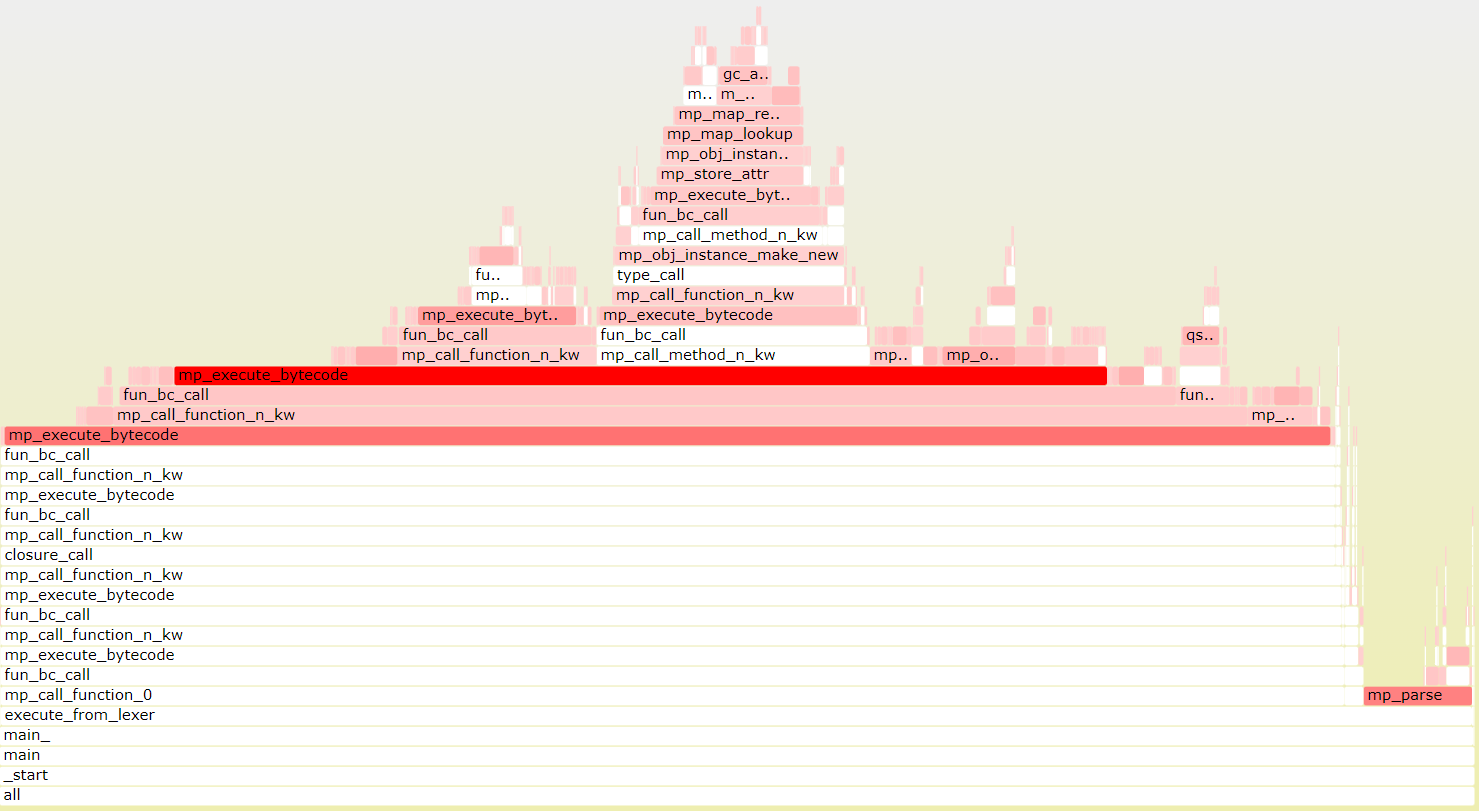}
	  \Description{Differential flame graph, showing mp\_execute\_bytecode() in deep red, mp\_parse() in medium red, and all other functions in light red or white.}
  \end{center}
  \caption{\label{fig:flame-dstackfix} 
	Differential Flame Graph (instructions retired) of purecap interpreter (after \texttt{VM\_MAX\_STATE\_ON\_STACK} adjustment) running \texttt{pystone} benchmark, relative to hybrid.}
\end{figure*}

As the individual flame graphs in Figures~\ref{fig:flame-ainit} and \ref{fig:flame-cstackfix} are now similar enough to be difficult to distinguish by eye, Figure~\ref{fig:flame-dstackfix} shows the \textit{differential} flame graph, where the difference in sample counts within a given function is indicated by the colour of that function's box. Red indicates an overhead (purecap $>$ hybrid) while blue indicates a saving (purecap $<$ hybrid); the saturation in either case indicates the magnitude of the difference.

From this graph we can see that the main remaining overhead (in instructions retired, at least) is localised in the function \verb`mp_execute_bytecode()`. This is the function that implements the MicroPython VM, interpreting the MPY bytecode for a given function by means of a computed-goto statement utilising a branch table, with each branch handling a particular bytecode instruction and then (unless execution halts due to a return or exception) performing another computed-goto to process the next instruction. 

\begin{lstlisting}[float=bht,label=lst:dispatch,caption={A64 (left) and C64 (right) disassembly for the \texttt{DISPATCH} sequence.}] 
; x23/c21 : instruction pointer (ip).
; x25/c27 : pointer to code_state struct
; [...,#0]: ip value before fetching the
            current bytecode instruction
; x26/c28 : pointer to branch table
str	x23, [x25]	str c21,[c27,#0]
ldrb	w8, [x23]	ldrb	w8, [c21]
add	x9, x23, #1	add	c1, c21, #1
mov	x28, x23	mov	c20, c21
mov	x23, x9		mov	c21, c1
ldr	x8, \		ldr	c0, \
   [x26,x8,lsl #3]         [c28,x8,lsl #4]
br	x8		br	c0
\end{lstlisting}

This is functionally a very tight loop -- the \verb`DISPATCH` sequence (saving the instruction pointer for exception-handling purposes and then performing the computed-goto) takes seven machine instructions (as shown in Listing~\ref{lst:dispatch}). For the simpler bytecode instructions, the actual execution only takes one to four machine instructions, for a total of eight to eleven machine instructions per bytecode instruction.

While the \verb`DISPATCH` sequence is compiled to the same number of machine instructions on both hybrid and purecap builds (seven), the same is not true for the other statements in the loop: there are several places where the purecap build uses more machine instructions to execute a given bytecode instruction than the hybrid build. As several of the bytecode instructions with a high ratio of capability instructions (known as C64) to generic Arm instructions (known as A64) were used far more frequently in \verb`pystone` than other benchmarks, we conjectured that this was the cause of the overhead we saw.

\begin{lstlisting}[float=ht,label=lst:ineff,caption={A64 (left) and C64 (right) disassembly of selected areas highlighting compiler inefficiency. Occurrence counts ignore the particular registers and values used.}] 
; Loading a constant uintptr_t
; (9 occurrences using ADD, 1 using SUB)
		mov x0, xzr
mov w8, #14	add c0, c0, #14

; Casting an integer to a uintptr_t
; (4 occurrences)
		mov x0, xzr
<N/A>		add c0, c0, x8, uxtx

; Comparing a uintptr_t against a constant
; (3 occurrences)
		mov x0, xzr
		add c0, c0, #6
cmp x22, #6	cmp x24, x0

; Testing a low-bits "tag" on a pointer
; (2 occurrences)
		and x8, x0, #0x2
		scvalue	c0, c0, x8
tbz w8, #1, ...	cbz x0, ...

; Load/store pairs of registers
; (4 occurrences using LDP, 2 using STP)
		ldur c0, ...
ldp x0, x1, ...	ldr c1, ...
\end{lstlisting}

While some of these extra machine instructions are unavoidable (for example, `tagging' the low bits of a hybrid pointer requires a single \verb`ORR` instruction while the same operation on a capability requires \verb`ORR` followed by \verb`SCVALUE`), a significant number of these are due to what appear to be odd choices by the compiler code generator. Listing~\ref{lst:ineff} shows some of the more obvious cases of inefficiency. In the first two cases, the compiler generates extra instructions seemingly to ensure the upper bits of the capability are cleared. However, since writing to the 32- or 64-bit view of a capability register is defined to clear the capability metadata and tag,\cite{morello2021} these `cast' instructions are unnecessary. The third case takes this even further, with the compiler generating a `cast' of the constant to a capability before performing an X-register
(i.e.\ non-capability) compare. In the fourth case, extra instructions are used to derive a new capability with a bit-masked value when the bit mask was only needed to test whether a particular bit was set. The final case involves breaking \verb`LDP/STP` instructions into pairs of single-register loads and stores. This notably did not happen everywhere: in many places the purecap binary did contain \verb`LDP/STP` instructions.

\begin{figure*}[ht]
  \begin{center}
    \includegraphics[width=\textwidth]{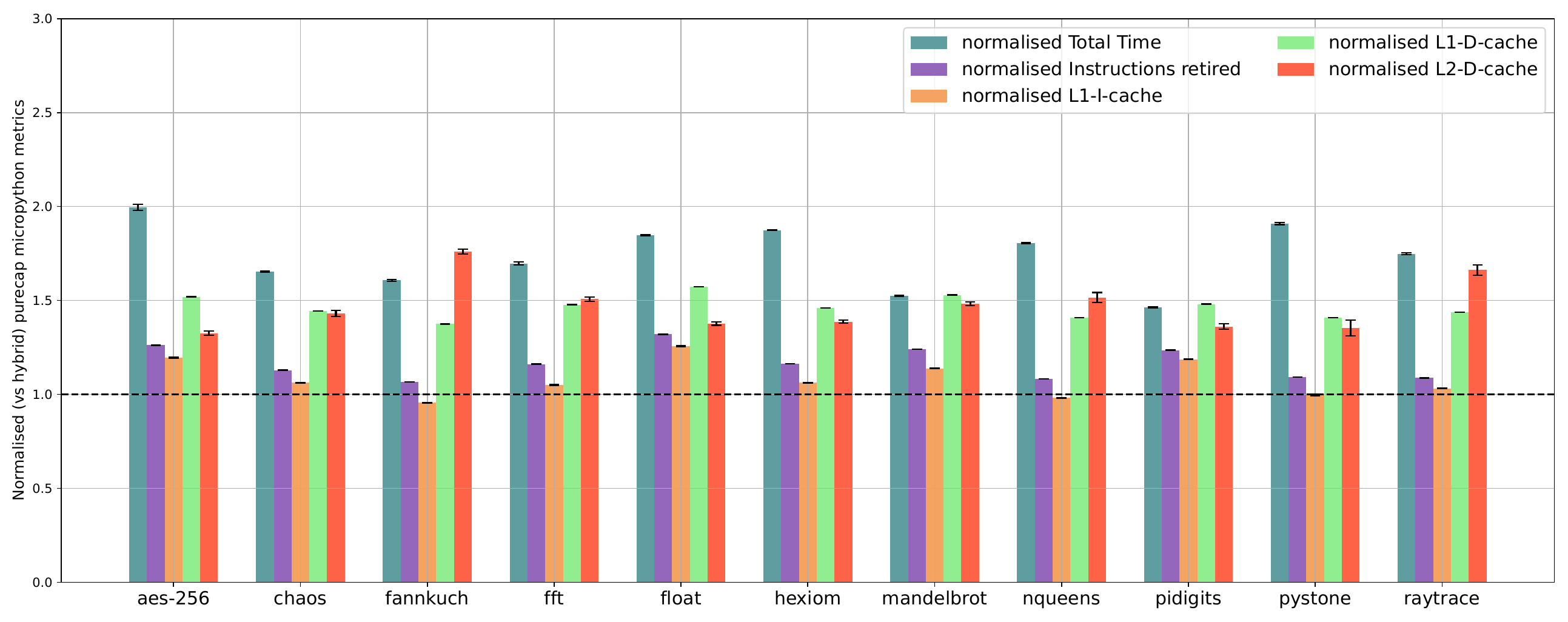}
	  \Description{A bar chart showing purecap MicroPython performance metrics, normalised against hybrid performance, for eleven benchmarks. Normalised total time varies between $1.5$ and $2.0$ for most benchmarks; other metrics vary between $1.0$ and $1.5$ for most benchmarks.}
  \end{center}
  \caption{\label{fig:perf_eval_asm} Performance of Python benchmarks running on the purecap interpreter with increased block size and maximum VM state size and a patched binary eliminating certain compiler inefficiencies, normalised to the hybrid interpreter performance.}
\end{figure*}

While these inefficiencies may seem insignificant at first glance, in a `hot' loop of often fewer than 20 instructions in length they quickly add up. Unfortunately, these issues cannot be cleanly fixed in the MicroPython source but rather require changes to the C compiler. In order to determine the actual impact of these inefficiencies (and as a change to the C compiler was out of scope for the project), we patched the binary by hand to fix several of these inefficiencies, saving a total of 19 machine instructions across the various branches. The normalised performance of the purecap build after this patch is shown in Figure~\ref{fig:perf_eval_asm}. The normalised execution time of \verb`pystone` has fallen from 2.8 to 1.9, and the normalised instructions-retired count has fallen to a similar range as the other benchmarks. It should be stressed that we have by no means eliminated all of the compiler-introduced inefficiencies, and as the \verb`llvm-morello` toolchain and other CHERI-aware compilers improve, we are likely to see significantly better performance from purecap software. 

\section{Conclusion}
\label{sec:concl}

When porting systems software to Morello or other CHERI platforms, the focus is often on the \textit{correctness} of the port. 
The issues identified in Sections~\ref{sec:block} and \ref{sec:frame} show that this is not sufficient: where parameters have been tuned in the original code based on an expectation of equal integer and pointer size, a CHERI port which does not adjust these parameters accordingly may exhibit unacceptable performance overheads despite being functionally correct.



After implementing the performance fixes described in this paper, all of our benchmarks showed between 1.4x and 2.0x execution time overhead on purecap, while overheads in terms of instructions retired ranged between 6.6\% and 32\% (Table~\ref{tab:perf_final}). The results in Section~\ref{sec:asm} suggest that a significant amount of this overhead is due to inefficiencies in the compiler. The performance results here are also likely to be further mitigated as the new hardware is released that is better tuned for CHERI operations than the current Morello
\cite{UCAM-CL-TR-986}.

\begin{table}[h]
 \caption{Performance overheads on purecap relative to hybrid by metric. Shown are the benchmark with the lowest overhead, the benchmark with the highest overhead, and the geometric mean across all the benchmarks. }
 \label{tab:perf_final}
  \centering
 \begin{tabular}{c c c c}
  \toprule
  Metric	      & Best   & Worst  & Geometric mean \\
  \midrule
  CPU cycles          & 45.4\% & 87.0\% & 64.8\% \\
  Instructions retired&  6.6\% & 32.1\% & 16.4\% \\
  Cycles/instruction  & 17.8\% & 71.4\% & 41.7\% \\
  L1-D-cache          & 37.5\% & 57.3\% & 46.3\% \\
  L1-I-cache          & -4.3\% & 25.7\% &  8.0\% \\
  L2-D-cache          & 32.6\% & 76.1\% & 46.4\% \\
  Total time          & 46.3\% & 99.6\% & 73.1\% \\
  \bottomrule
 \end{tabular}
\end{table}

\begin{acks}
This work was funded by the Digital Security by Design (DSbD) programme delivered by UKRI
(including grants EP/V000349/1 and EP/X015831/1),
also by the UK Defence and Security Accelerator contract ACC6037520.
\end{acks}

\balance
\bibliographystyle{ACM-Reference-Format}
\bibliography{paper}

\end{document}